Thermoelectric prospects of nanomaterials with spin-orbit surface bands.

T. E. Huber,[1*] K. Owusu,[1] S. Johnson,[2] A. Nikolaeva[3,4] L. Konopko,[3] R. C. Johnson[5], and M. J. Graf [5]

[1] Howard University, Washington, DC 20059

[2] Prince Georges Community College, Largo, MD 20774

[2] Academy of Sciences, Chisinau, Moldova

[3] International Laboratory of High Magnetic Fields and Low Temperatures, Wroclaw, Poland.

[5] Department of Physics, Boston College, Chestnut Hill, MA 02467

Nanostructured composites and nanowire arrays of traditional thermoelectrics like Bi, $Bi_{1-x}Sb_x$ and $Bi_2Te_3$ have metallic Rashba surface spin-orbit bands featuring high mobilities rivaling that of the bulk for which topological insulator behavior has been proposed. Nearly pure surface electronic transport has been observed at low temperatures in Bi nanowires with diameter around the critical diameter, 50 nm, for the semimetal-to semiconductor transition. The surface contributes strongly to the thermopower, actually dominating for temperatures $T < 100$ K in these nanowires. The surface thermopower was found to be $-1\ T\ \mu V/K^2$, a value that is consistent with theory. We show that surface electronic transport together with boundary phonon scattering leads to enhanced thermoelectric performance at low temperatures of Bi nanowire arrays. We compare with bulk n-BiSb alloys, optimized $CsBi_4Te_6$ and optimized $Bi_2Te_3$. Surface dominated electronic transport can be expected in nanomaterials of the other traditional thermoelectrics.

1. INTRODUCTION

Materials of high thermoelectric (TE) figure of merit $zT = T\alpha^2\sigma/\kappa$ where $T$ is the absolute temperature, $\alpha$ is the thermopower, $\sigma$ is the electrical conductivity and $\kappa$ the total thermal conductivity, are employed in solid state cooling devices and may be used in future waste heat recovery and heat energy harvesting applications.[1] Bismuth, $Bi_{(1-x)}Sb_x$ (x~0.1) and $Bi_2Te_3$ are the traditional thermoelectric (TE) materials for low and room-temperature operation that have $zT$ ~ 1. The material science of nanostructured TEs has advanced substantially in the past decade. Gains as high as a factor of two have been demonstrated with nanostructured $Bi_2Te_3$ consisting of arrays of 1-5 nm diameter particles.[2] Such gains in $zT$ are commonly associated with phonon and electronic phenomena in nanostructured materials. Phonon-boundary scattering decrease the phonon contribution to $\kappa$ tending to increase $zT$.[3] The electronic properties of the bulklike bands, $\alpha$ and $\sigma$, are modified also. In conduits, such as nanowires (NW) and constrictions in composites, characterized by a diameter $d$, electronic finite size effects[4] (arising when the electronic mean free path $mfp_e \sim d$, roughly) decrease $\sigma$ without large changes in $\alpha$ and therefore tend to decrease $zT$. However, further decreases in $d$ are considered to be beneficial because of quantum size effects that appear when the quantum confinement energy $E_c = \hbar^2\pi^2/2m^*m_e d^2$ (where $m^*$ is the bulk carrier effective mass and $m_e$ is the free electron mass) is sufficiently large to allow manipulation of carrier densities and doping.[5] However, the TE effects arising because surface charge conduction and phonons scattering at the boundaries have not been considered. Here we show that these factors change the outlook for TEs substantially.

The reason for the new surface band in TEs is spin-orbit interaction (SOI) at the surface. Traditional TEs are formed from heavy elements that have a potential for large SOI. However, in bulk solids, time reversal symmetry combined with space inversion symmetry depresses SOI. In same cases SOI is associated with gaps in the bulk bands. By contrast, at the crystal surface of semi-infinite surfaces of TEs, specifically bismuth, BiSb and $Bi_2Te_3$, space symmetry is lost and, at the surface, SOI effects are sufficiently strong that they give rise to a new band which is distinct from the bulklike band. Since this band arises because of SOI it is interesting for spintronic applications. Surface state bands were first observed spectroscopically with angle resolved photoemission spectroscopy (ARPES)[6] in Bi crystals and soon after were identified in electronic transport of Bi nanowires.[7] Surface band conduction is relevant in nanostructures because their high surface-to-volume ratio. Moreover, the recent discovery that selected bulk TEs like $Bi_2Te_3$ and $Bi_2Se_3$ feature exotic three dimensional (3D) topological insulator (TI) behavior[8] created new possibilities because for TIs the surface state would be protected from dissipation by time reversal symmetry and therefore have exotic spintronic properties and high mobility. Takahashi *et al*[9] and Ghaemi *et el*[10] presented models of the thermopower of surface states in interplay with bulklike carriers in thin films and found that surface dominates at low temperatures. They were motivated by the report of Hor *et al*[11] of enhancements in the thermopower of $Bi_2Se_3$; however, the surface origin of the thermopower in these experiments is uncertain. This is not surprising since significant experimental hurdles exist (as well as fundamental reasons) for realizing pure surface conduction; in the field of TIs it is observed that the most likely candidates according to theory, $Bi_2Te_3$ and $Bi_2Se_3$, are in fact not very good bulk insulators in the laboratory. Surface state band mobilities in $Bi_2Te_3$ are low[12] ($1 \times 10^4$ $cm^2 sec^{-1} V^{-1}$) and unobservable in $Bi_2Se_3$.[13] Bi is classified as a trivial topological insulator[14,15] which means that the surface states of Bi are not expected theoretically to be topologically protected from dissipation. Also there is substantial bulk-surface state hybridization in Bi for some crystalline orientations.[16] that may circumvent TI behavior. However, Ghaemi *et al* show that even in a strong TI like $Bi_2Se_3$ that has a single massless Dirac cone hybridization between the top and bottom surfaces in a nanostructure can cause the charge carriers to be massive and the surface bands to be gapped. Bi features very low intrinsic dissipation in comparison with the strong TIs and therefore is a good candidate to exhibit high surface mobilities.

Bi is a semimetal where the overlap energy $E_0$ ~37 meV between the electron and hole band leads to a significant electron *n* and hole *p* density ($n = p = n_0 = 3 \times 10^{17}$ cm$^{-3}$). The TE properties of Bi are well known.[17] Also, in nanowires, when the confinement energy $E_c > E_o$ for *d* ~ 50 nm the overlap becomes a gap and *n* and *p* can decrease critically below $n_0$; this phenomenon that is called the semimetal-to-semiconductor (SMSC) transition.[5,18] Huber, Adeyeye, Nikolaeva, Konopko, Johnson, and Graf[19] (HANKJG) studied electronic transport, resistance and thermopower, of 20-, 30-, 50- and 200-nm Bi nanowire arrays - that is, nanowires that were both on the semimetal and semiconductor side of the SMSC were studied. The trends that were observed are reminiscent to those observed by Boukai, Xu, and Heath[20] and Lin, Rabin, Cronin, Ying and Dresselhaus.[21] HANKJG also presented a Shubnikov-de Haas (SdH) study of the magnetoresistance oscillations caused by surface electrons and bulklike holes that enable determination of their Fermi surfaces, densities and mobilities. Surface electrons are in 3D Fermi surfaces. They showed that for *d*~ 50 nm the SMSC has the effect of decreasing the bulk population several fold from $n_0$, the bulk value. Surface electrons are not very modified by confinement and have a density $N = -1.3 \times 10^{13}$ cm$^{-2}$. In these nanowires, surface high mobilities exceeding 2 m$^2$sec$^{-1}$V$^{-1}$ are observed and contribute strongly to the thermopower, dominating for temperatures $T < 100$ K. The surface thermopower is $-1.2\ T$ µV/K$^2$, a value that is consistent with theory considering the density of surface states in Bi surfaces. For 50-nm NWs, for which the hole density is the smallest among all the NWs that were investigated, the 50 K surface charges thermopower is - 60 µV/K². The bulklike holes contribute + 35 µV/K². Basing ourselves in these studies, here we carry out an evaluation of the expected *zT* in NW arrays and nanostructures of Bi and BiSb and present a discussion of the prospects of Rashba surface band-dominated nanostructured traditional TE materials.

## II. THERMOELECTRIC FIGURE OF MERIT

The diffusive thermopower (*mfp* < wirelength) is given by: [22]

$$\alpha_\Sigma = \frac{(k_B^2 \pi^2 T / 3e)}{E_F^\Sigma} \left[ r + (d \ln N / d \ln E)_{E_F^\Sigma} \right] \quad (1)$$

Here $E_F^\Sigma$ is the surface band Fermi energy, $r = (\partial \ln \tau / \partial \ln E)_{E_F^\Sigma}$ where $\tau$ is the carrier lifetime. We assume $r \sim 0$, which is appropriate in this case as the lifetime is dominated by boundary scattering. $(\partial \ln N / \partial \ln E)_{E_F^\Sigma} = 3/2$ in 3D. From $N = -1.3 \times 10^{13}$ cm⁻², the Fermi energy is 18 meV and, from Equation (1), we find $\alpha_\Sigma = -1.2\ T$ µV/K² where the sign of the partial thermopower is related unambiguously to the sign of the charge of the carriers. In comparison, the low temperature thermopowers of electrons $\alpha_e$ and holes $\alpha_h$ in bulk Bi are found to be approximately $-1\ T$ µV/K² and $+3\ T$ µV/K² respectively.[17] In Bi nanowires, out-of-equilibrium effects like the phonon-drag effect are of negligible importance since phonon scattering is mainly phonon-boundary rather than phonon-carrier.[23] This model for decreased phonon drag has been observed to apply in other cases of thermopower of nanowires, notably sub-100 nm diameter Si point contacts,[24] and Si and Ni nanowires, respectively.[25,26] This is relevant because Equation (1) is not applicable in the case that there are phonon drag effects and also because such non-equilibrium effects limit $zT$ to values that are much less than one.[1]

The normalized resistance and thermopower of the nanowires are shown in Figure 1.a and Figure 2, respectively. HANKJG argued that the saturation of the resistance at low $T$ in the normalized resistance and the trend toward negative thermopower for decreasing $d$ can be associated with type-n surface states in Bi nanowires. The reason that surface states become dominant is that the surface to volume ratio increases as $1/d$ and the SMSC transition makes the bulk in the interior of the wires a semiconductor.

The data that is presented by HANKJG allows direct calculation of $\alpha$ and $\sigma$. The TE figure of merit $zT$ has not been measured directly but can be obtained from $T\alpha^2\sigma/\kappa$. However, the thermal conductivity was not measured directly. The (total) $\kappa$ is $\kappa = \kappa_{electr} + \kappa_{phonon}$, where the first and second terms are the electronic and phonon conductivity, respectively. A. Moore, M.T. Pettes, F. Zhou, and L. Shi presented a study of the $\kappa$ of individual Bi nanowires with diameter in the range between 280 and 170 nm in the temperature range between 100 K and 300 K.[27] We show Moore's data in Fig. 1.b. A trend of decreasing $\kappa$ for decreasing diameter was observed; this is expected in terms of electronic and phonon size effects. Therefore there is no reason to expect that this trend will reverse for finer nanowires and therefore, the thermal conductivity in 30 nm and 20 nm nanowires can be expected to be even less than that for 150 nm nanowires. Measurements of $\kappa$ of Bi composites by Song, Shen, Dunn, Moore, Goorsky, Radetic, Gronsky, and Chen that include fine, $d$ down to 20 nm, nanocomposites support this expectation.[28]

Another estimate of the thermal conductivity relates to the electronic part. $\kappa_{electr}$ can be estimated from the electronic conductivity $\sigma$ assuming the very general Wiedeman-Franz (WF) law:

$$\kappa_{electr} = L_0 \sigma(T) T \qquad (2)$$

where $L_0 = \frac{\pi^2}{3}(\frac{k_B}{e})^2 = 2.4 \times 10^{-8}$ (Volt)$^2$/K$^2$ is the usual Lorenz number. We are assuming that the electron gas is degenerate, a simplifying approximation.[1,29] Since the normalized resistance is known, a close (better than order of magnitude) estimate of the absolute conductivity can be obtained by considering that the room temperature resistance $R$ (300 K) is more than $\rho_0 L/(\pi/4)d^2$, where $\rho_0$ is the 300 K resistivity.[13] This is a close inequality because finite size effects are small and positive at 300 K, which is expected theoretically because the *mfp* is short at room temperatures. The weakness of finite size effects at room temperatures is observed in four-point measurements of isolated individual Bi nanowires.[20] It is also observed in another work where very long Bi NWs, prepared by the Ulitovsky method,[30] were employed. This fabrication method employs glass-pulling techniques similar to those used in glass fiber-making to produce a single strand of Bi NW of controllable diameter in a glass fiber. In this work, the measurement is performed using two contacts; because the length of the NW employed in this study is very long, a fraction of a millimeter, the resistance of the NWs is high, and contact resistance at the end of the NWs is within manageable limits. W*e* show our estimate of $\kappa_{electr}$ in the case of our nanowires in Figure 1.b and we find an inconsistency in that the expected electronic term is larger than the measured total thermal conductivity. However, in these experiments electrical contact was not made to the nanowires and therefore electrical conductivity was not measured. The $\kappa_{electr}$ term depends upon $\sigma$ that, in turn, exhibits great variability depending upon fabrication and preparation. One way to reconcile the two measurements is to assume that the actual electrical conductivity in Moore case was much smaller than in HANKJG case and the phonon thermal conductivity is much smaller than the electrical thermal conductivity.

If $\kappa_{electr} \gg \kappa_{phonon}$, that is for composites where the phonon conductivity is completely quenched, we have

$$zT = \frac{\alpha^2}{(\kappa/\sigma)}T = \frac{\alpha^2}{L_0} \tag{3}$$

In their analysis of bulk Bi thermoelectricity, Gallo et al remarked that, since the macroscopic parameter Z is a sensitive function of the carrier concentrations, it is convenient to speak of a "hypothetical optimum index of efficiency," $z_oT$. $z_o$ occurs only when electronic transport involves only one type of carrier (electrons), ensuring (a) that the Seebeck coefficient is a maximum because holes do not participate and (b) that thermal conduction by bipolar diffusion is essentially zero. The hypothetical optimum applies to the case of a hypothetical Bi alloy whereby the electrons dominate absolutely (we do not know that such an alloy exists). We introduce the same concept for Bi nanostructures. We use Equation 3 to arrive at the hypothetical optimum values for nanostructures where only surface states contribute, shown in Fig. 3. Inspection of Figure 2 shows that the most favorable case is the one of 50-nm Bi nanowires[19] that exhibits pure surface conduction for $T < \sim 50$ K. For higher temperatures, holes contribute to electronic transport and decrease $\alpha$ and $zT$. In this regard, the study by Lin et al of 45-nm BiSb nanowires is interesting in that their thermopower is linear for temperatures as high as 100 K.[21] We have compared nanowire arrays to the select number of materials that are known to have large $zT$ at low temperature. These results are shown in Figure 3 also. Most successful are the $Bi_{1-x}Sb_x$ alloys. We show representative data from Redko.[31] This data shares many commonalities with the data presented by Lenoir et al.[32] The data labeled with $Bi_2Te_3$ is representative of the (Bi-Sb)$_2$(Te-Se)$_3$ alloys.[1] We also show data for CsBi$_4$Te$_6$.[33] The values of optimal hypothetical $zT$ of surface states and of 50-nm Bi and of 45-nm Bi(0.95)Sb(0.05) exceed that of the other materials that are known to display excellent TE properties at low temperatures. This capability can be employed directly in thermoelectric coolers where the fractional temperature decrease $(T_H - T_C)/T_C$ per stage is $(1/2)zT_C$.[1]

In practice, nanowire arrays are embedded in an alumina matrix that is amorphous. The $\kappa$ of the alumina at room temperature has been estimated[34] to be 1.4 W/(m K) which represents a small fraction of the $\kappa_{electr}$. Therefore we expect that the contribution of the matrix can be neglected in nanowire arrays.

## 4. CONCLUSION. PROSPECT OF NANOSTRUCTURED TRADITIONAL THERMOELECTRICS

The theory of topological insulators predicts many interesting properties, however substantial hurdles exist for realizing these predictions since the candidate materials are in fact poor bulk insulators. Our proposal for aiding bulk insulator behavior in semimetal like Bi, Sb, and BiSb is to shape these TEs into nanowires that become semiconductors as a result of bulk quantum confinement. We tested Bi nanowires that show strong quantum confinement effects for large diameters (comparatively to most of the other thermoelectrics) due to small bulk effective mass. 50-nm wires feature almost pure surface conduction. Bi is not a true topological insulator. Still, nanowires of Bi are exceptional materials that exhibit carrier mobilities of over 2 m$^2$sec$^{-1}$V$^{-1}$ with a density 2.2 $\times 10^{12}$ cm$^{-2}$. This mobility is large, 2/3 of the values that are found for unsuspended graphene,[35] with significantly lower charge densities. Mobility values are twice those found by Qu *et al*[12] for Bi$_2$Te$_3$ surface bands. The high value of surface mobility appear to be related to the special conditions in 50-nm wires since it is significantly less for 30- and 20-nm wires. 50-nm Bi nanowires are the most compelling case of pure surface conduction for $T < \sim 50$ K. As put forth by Lin *et al*[21] in their study of 45-nm BiSb nanowires, maybe the advantages of alloying and quantum confinement can be advantageously combined in future experiments with BiSb nanostructures.

Our estimate of the TE figure of merit is shown in Figure 3. The values of optimal hypothetical $zT$ of surface states and of 50-nm Bi and of 45-nm Bi(0.95)Sb(0.05) exceed that of the other materials that are known to display excellent TE properties at low temperatures. From this study it appears that nanowire arrays and composites based on nanowires of traditional TEs Bi, BiSb and also of other TEs that exhibit topological insulator behavior will be of practical interest for cooling and energy sensing. Since the surface bands arise because of spin-orbit interactions, spin dependent transport devices can probably be engineered using these materials and therefore spintronic applications may be realized in the future.

Many experimental studies of fine nanowires point out the relevance of phonon boundary scattering that results in a decreased phonon thermal conductivity. Unfortunately experiments have not measured all the properties of Bi (or other TIs) in the same system. HANKJG measured thermopower and electrical conductivity only, and not $\kappa$. Others, notably Moore *et al*[27] and Song *et al*[28] measured $\kappa$ but did not measure $\sigma$ or $\alpha$. Integrated experiments with state-of-the-art materials are necessary to fully understand the new TE nanomaterials.

This material is based upon work supported by the National Science Foundation (NSF) under Grant No. NSF-DMR-0839955 and NSF-DMR-0611595 and by the U.S. Army Research Office Materials Science Division under Grant No. W911NF-09-1-05-29. We also acknowledge support by the Boeing Corporation RA-6, Swiss National Science Foundation SCOPES and

Science Technology Center in the Ukraine-project N 5050,

# REFERENCES

*Corresponding author: T.E. Huber. Address: Howard University, Washington, DC 20059, USA. Tel: 202 806 6768. *Email address*: **thuber@howard.edu**.

1. H. J. Goldsmid in "Electronic Refrigeration" (Pion Limited, London, 1986).

2. R. Venkatasubramanian, E. Siivola, T. Colpitts and B. O'Quinn. Nature, **413**, 597 (2001).

3. D. G. Cahill, W. K. Ford, K. E. Goodson, G. D. Mahan, H. J. Maris, A. Majumdar, R. Merlin, S. R. Phillpot, Journal of Applied Physics, **93**, 793 (2003).

4. E. H. Sondheimer, Adv. Phys. **1**, 1 (1952).

5. L. D. Hicks and M. S. Dresselhaus, Phys. Rev. B **47**, 16631 (1993).

6. Ph. Hofmann, Prog. Surf. Sci. **81**, 191 (2006).

7. T. E. Huber, A. Nikolaeva, D. Gitsu, L. Konopko, C.A. Foss, Jr., and M.J. Graf. Applied Physics Letters **84**, 1326 (2004).

8. M. Z. Hasan and C. L. Kane, Rev. Mod. Phys. **82**, 3045 (2010).

9. R. Takahashi and S. Murakami, Phys. Rev. B **81**, 161302(R) (2010)

10. P. Ghaemi, R.S. K. Mong and J. E. Moore, Phys. Rev. Lett. **105**, 166603 (2010).

11. Y. S. Hor, A. Richardella, P. Roushan, Y. Xia, J. G. Checkelsky, A. Yazdani, M. Z. Hasan, N. P. Ong and R. J. Cava, Phys. Rev. B**79**, 195208 (2009).

12. D-X Qu, Y. S. Hor, J. Xiong, R. J. Cava and N. P. Ong, Science **329**, 821 (2010)

13. N. P. Butch, K. Kirshenbaum, P. Syers, A. B. Shuskov, G. S. Jenkins, H. D. Drew, and J. Paglione, Phys. Rev. B **81**, 241301(R) (2010).

14. Y. Liu and R. E. Allen, Phys. Rev. B **52**, 1566 (1995).

15. L. Fu and C. L. Kane, Phys. Rev. B **76**, 045302

16. T. Hirahara, T. Nagao, I. Matsuda, G. Bihlmayer, E. V. Chulkov, Yu. M. Korotev, P. M. Echenique, M. Saito and S. Hasegawa, Phys. Rev. Lett. **97** 146803 (2006).

17. C. F. Gallo, B. S. Chandrasekhar, and P. H. Sutter, J. Appl. Phys. **34**, 144 (1963).

18. T. E. Huber, A. Nikolaeva, L. Konopko, and M. J. Graf, Phys. Rev. B **79**, 201304-8 (R) (2009).

19. T. E. Huber, A. Adeyeye, A. Nikolaeva, L. Konopko, R. C. Johnson, and M. J. Graf. Phys. Rev. B**83**, 235414 (2011).

20. A. Boukai, K. Xu, and J. R. Heath. Adv. Mat. **18** 864 (2006)

21. Y-M Lin, D. Rabin, S. B. Cronin, J.Y. Ying and M. S. Dresselhaus. App. Phys. Lett. **81**, 2403 (2002).

22. R. Fletcher, Semicond. Sci. Technol. **14**, R1 (1999).

23. D. Gitsu, T. Huber, L. Konopko, and A. Nikolaeva. Phys. Stat. Sol. (b) **242**, 2497 (2005).

24. R.Trzcinski, E. Gmelin, and H. J. Queisser, Phys. Rev. Lett. **56**, 1086 (1986).

25. A. I. Boukai, Y. Bunimovich, J. Tahir-Kheli, J-K Yu, W.A. Goddard III and J.R. Heath. Nature **451**, 168 (2008).

26. A. Soni and G. S. Okram. Appl. Phys. Lett. **95**, 013101 (2009).

27. A. L. Moore, M.T. Pettes, F. Zhou, and L. Shi. J. Appl. Phys. **106**, 034310 (2009).

28. D. W. Song, W.N. Shen, B. Dunn, C.D. Moore, M. S. Goorsky, T. Radetic, R. Gronsky, and G. Chen. Appl. Phys. Lett. **84**, 1883 (2003).

29. N. W. Ashcroft and N.D. Mermin in "Solid State Physics" (Saunders College Publishing, Fort Worth, 1976), p 258.

30. A. Nikolaeva, T. E. Huber, D. Gitsu, and L. Konopko, Phys. Rev. B **77**, 035422 (2008).

31. N. A. Redko, Proceedings of the 14$^{th}$ International Conference on Thermoelectrics, (IEEE, Piscataway, NJ. USA, 1995).

32. B. Lenoir, M. Cassart, J-P Michenaud, H. Scherrer, and S. Scherrer. J. Phys. Chem Solids. **57**, 89 (1996).

33. D. Chung, T. Hogan, P. Bazis, M. Tocci-Lane, C. Kannenwurf, M. Bastea, C. Uher, and M. Kanatzidis. Science **287**, 1024 (2000).

34. D-A Borca-Tasciuc, G. Chen, G. Chen, A. Prieto, M. S. Martin-Gonzalez, A. Stacy, T. Sands, M. A. Ryan and J-P Fleurial, Appl. Phys. Lett. **85**, 6001 (2004).

35. K. I. Bolodin, K. J. Sikes, Z. Jiang, M. Klima, G. Fudenberg, J. Hone, P. Kim and H. L. Stormer, Solid State Comm. **146** 351 (2008).

# FIGURE CAPTIONS

**FIGURE 1.** (a) Inset. SEM image of the top of a 50-nm Bi nanowire array. Light spots represent nanowires. Electron energy is 10 keV. Normalized resistance of arrays of 200-, 50-, 30-, and 20-nm Bi nanowires as indicated. (b) Thermal conductivity from various sources as indicated. Solid line: $\kappa_{electr}$ evaluated using Equation 2. Dashed lines: $\kappa_{phonon}$ from Moore, Pettes, Zhou and Shi,[27] diameters are indicated.

**FIGURE 2.** Inset: Anvil-type experimental set-up for thermopower $\alpha$ of massive nanowire arrays. The set-up assumes that there is a heater that maintains heat current from the hot (H) to the cold (C) side. Then $\alpha = (V_H - V_C)/(T_H - T_C)$ where $V_H$ and $V_C$ are the electrochemical potentials. $T_H$ $T_C$. Main panel: Thermopower of 200-, 50-, 30-, and 20-nm Bi nanowires as indicated. 40-nm data from Lin, Rabin, Cronin, Ying and Dresselhaus[21] is also shown. The dashed line on the 50-nm data is a linear fit.

**FIGURE 3.** Thermoelectric figure of merit $zT$. Hypothetical optimum $zT$ of arrays of nanowires featuring surface state bands in comparison to that of bulk n-BiSb,[32] optimized $CsBi_4Te_6$ (Ref. 33) and optimized $Bi_2Te_3$ (Ref. 1).

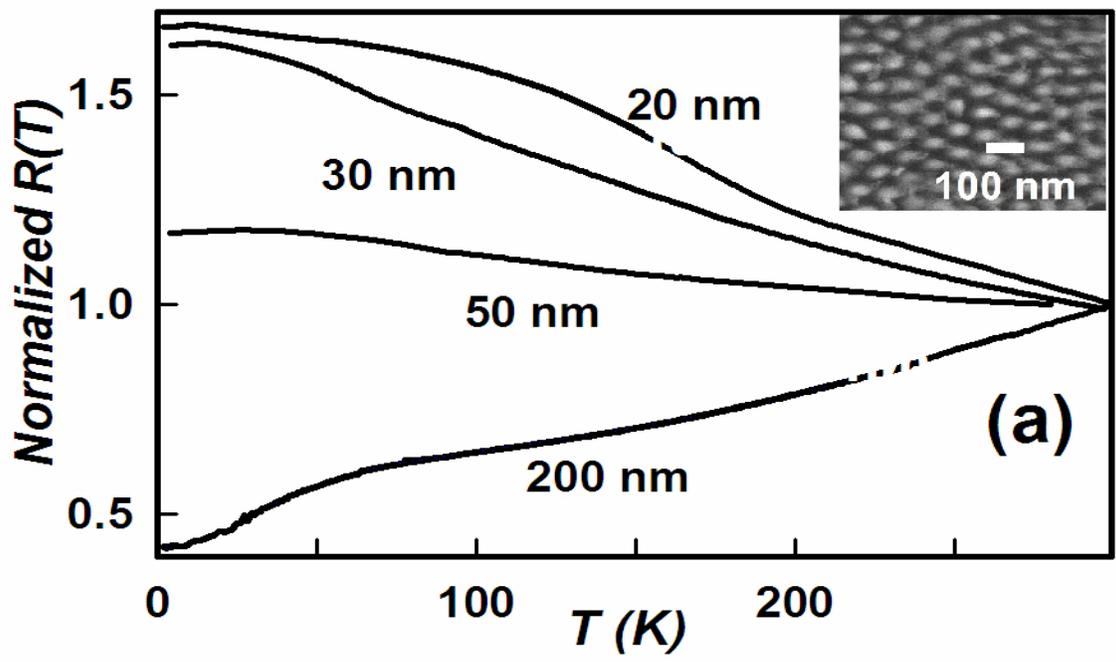

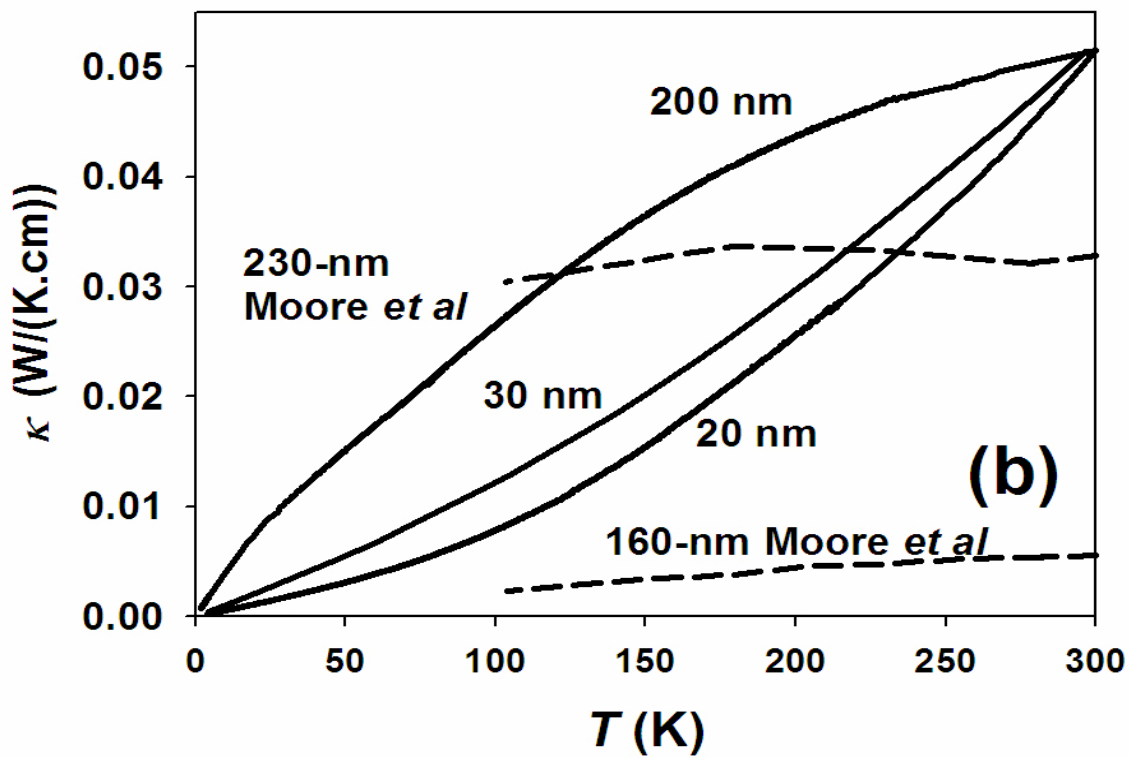

Figure 1a and b. Huber *et al.* (2011).

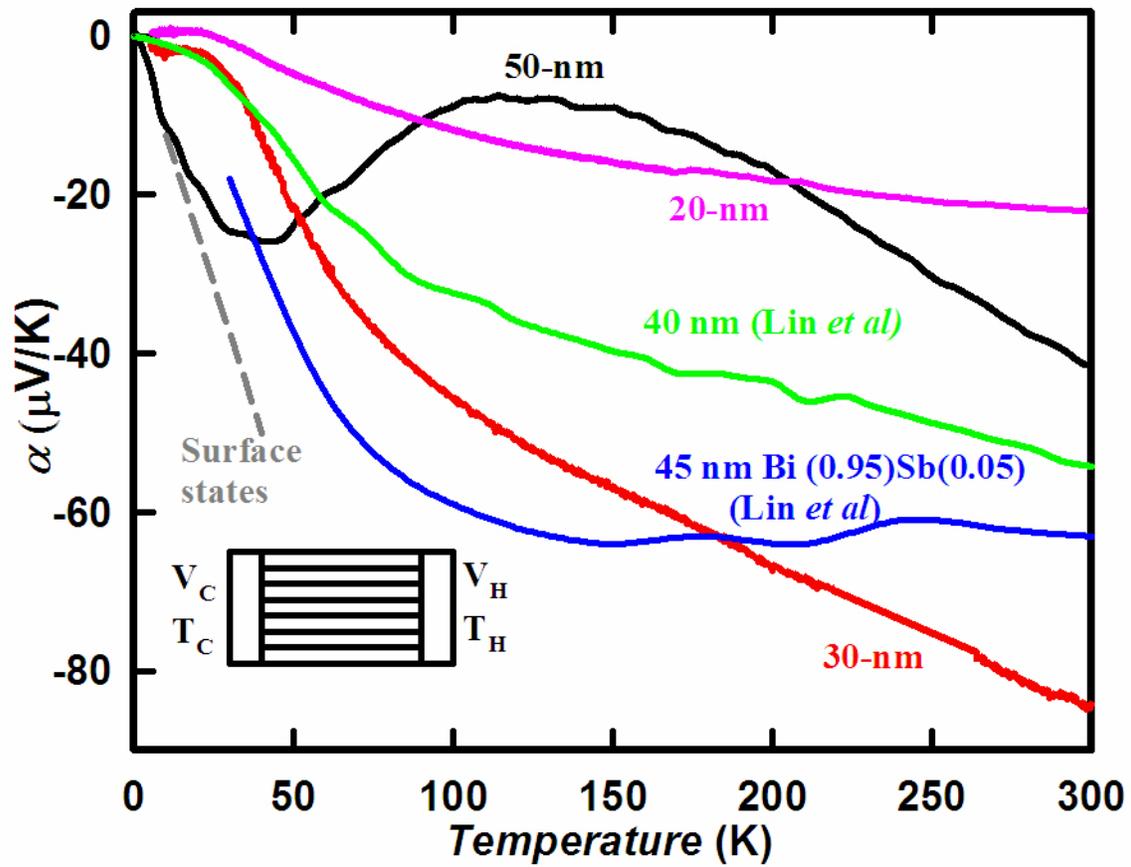

Figure 2. Huber et al. (2011).

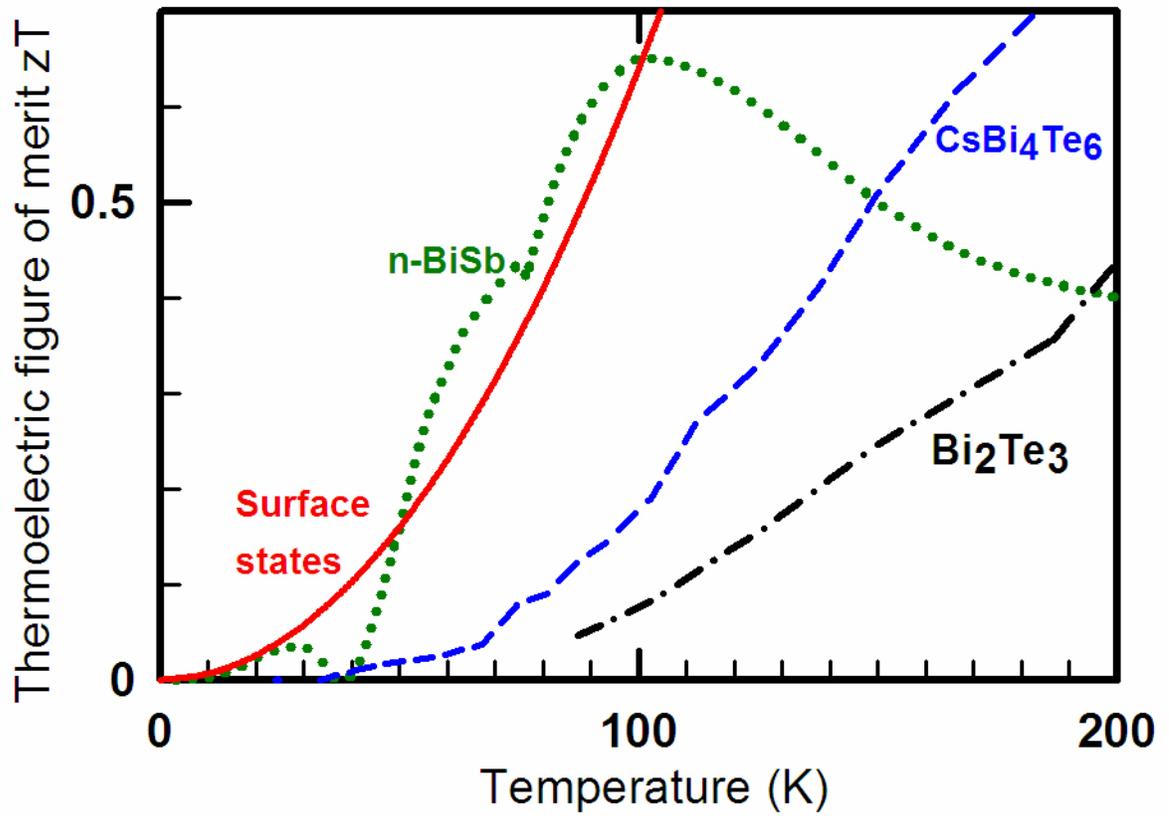

Figure 3. Huber et al (2011).